\documentclass[5p,final]{elsarticle}
\usepackage[utf8]{inputenc}
\usepackage[T1]{fontenc}
\usepackage{lmodern}
\usepackage{amsmath,amssymb,amsfonts,amsthm}
\usepackage{graphicx}
\usepackage{bm}
\usepackage{microtype}
\makeatletter
\def\ps@pprintTitle{%
  \let\@oddhead\@empty
  \let\@evenhead\@empty
  \let\@oddfoot\@empty
  \let\@evenfoot\@empty}
\makeatother
\usepackage[colorlinks=true,allcolors=blue]{hyperref}
\usepackage{natbib}
\providecommand{\citea}[1]{\cite{#1}}
\newcommand{\E}{\mathbb{E}}
\newcommand{\Pp}{\mathbb{P}}

\begin{document}
\begin{frontmatter}
\title{Event-Time Order-Flow Memory, Operational-Time Impact, and Subordinated Market Observables}
\author[unsw-mathstats]{Christopher Angstmann}
\address[unsw-mathstats]{School of Mathematics and Statistics, University of New South Wales, Sydney, NSW 2052, Australia}
\author[uct-sta]{Tim Gebbie}
\address[uct-sta]{Department of Statistical Sciences, University of Cape Town, Rondebosch 7701, Western Cape, South Africa}
\begin{abstract}
We consider two canonical market-microstructure regularities: the long-memory of trade signs and the square-root law of meta-order impact. The point is not to propose new empirical laws, but to separate the clocks on which existing laws are defined. The sign-memory law is an event-time statement about the ordering and fragmentation of hidden orders. The square-root impact law is an operational-time statement about front motion in a locally linear latent order book. Starting from a discrete-time random-walk bid/ask reaction--diffusion order book with a separate event clock, we identify an event-time imbalance reduction, derive the operational-time front and impact equations in the locally linear regime, and then subordinate both sign and impact observables to calendar time. Fractional or tempered clock effects enter through this event-to-calendar projection, not through a different operational-time impact mechanism. This gives a compact clock-aware framework in which event-time sign persistence, operational-time square-root impact, and anomalous calendar-time effects appear as distinct but compatible consequences of a common order-book representation.
\end{abstract}
\end{frontmatter}
\section{Introduction}

Two stylized facts dominate discussions of market microstructure. The first is the long-memory of child-order signs, associated with order splitting and meta-orders \cite{lillo2005theory,lillo2004efficient,sato2023prl,sato2024jsp}. The second is the concave, approximately square-root law of meta-order impact, which can be derived using latent-liquidity and reaction--diffusion descriptions of the order book \cite{toth2011anomalous,donier2015fully,mastromatteo2014latent,benzaquen2018multi}. For a broader review of order flow, price formation, market impact, and execution costs, see \cite{lillo2021orderflow}. These regularities are often presented as if they belonged to a common time variable \cite{angstmann2026nonunique}; here we argue that they are not. In contrast to continuous-time constructions \cite{toth2011anomalous,nadtochiy2022simple}, which explain concave meta-order impact through state-dependent visible-book or latent-book variables, the emphasis here is on the construction of time itself.

The central message here is that the long-memory law and the square-root law live on different clocks. The trade-sign law is an \emph{event-time} statement: it concerns the ordering of child orders and the persistence created when long hidden orders are split into many same-sign pieces \cite{lillo2005theory,sato2023prl}. The impact law is an \emph{operational-time} statement: it concerns the evolution of an imbalance field and its moving reaction front under diffusive transport, replenishment, and signed forcing \cite{toth2011anomalous,donier2015fully}. Calendar-time observables are then obtained by subordination through the waiting-time process between order-book events, in direct analogy with discrete-time random walk constructions used in anomalous diffusion \cite{angstmann2015dtrw,meerschaert2012fractional,diana2024anomalous}.

The novelty is the common embedding. The renewal law, front equation, and calendar-time observable are not combined by identifying their time variables. They are composed through an explicit hierarchy of event, operational, and calendar time. Different approximations apply at the three stages. The bid/ask imbalance cancellation is exact for the symmetric update used below. The Lillo--Mike--Farmer sign-memory relation is exact under the sequential hidden-order renewal assumptions. The square-root impact law is an operational-time asymptotic obtained under the locally linear, frozen-background latent-order-book approximation. The calendar-time statements are subordination statements conditional on the chosen event-clock model.

\begin{figure*}[t]
\centering
\includegraphics[width=\textwidth]{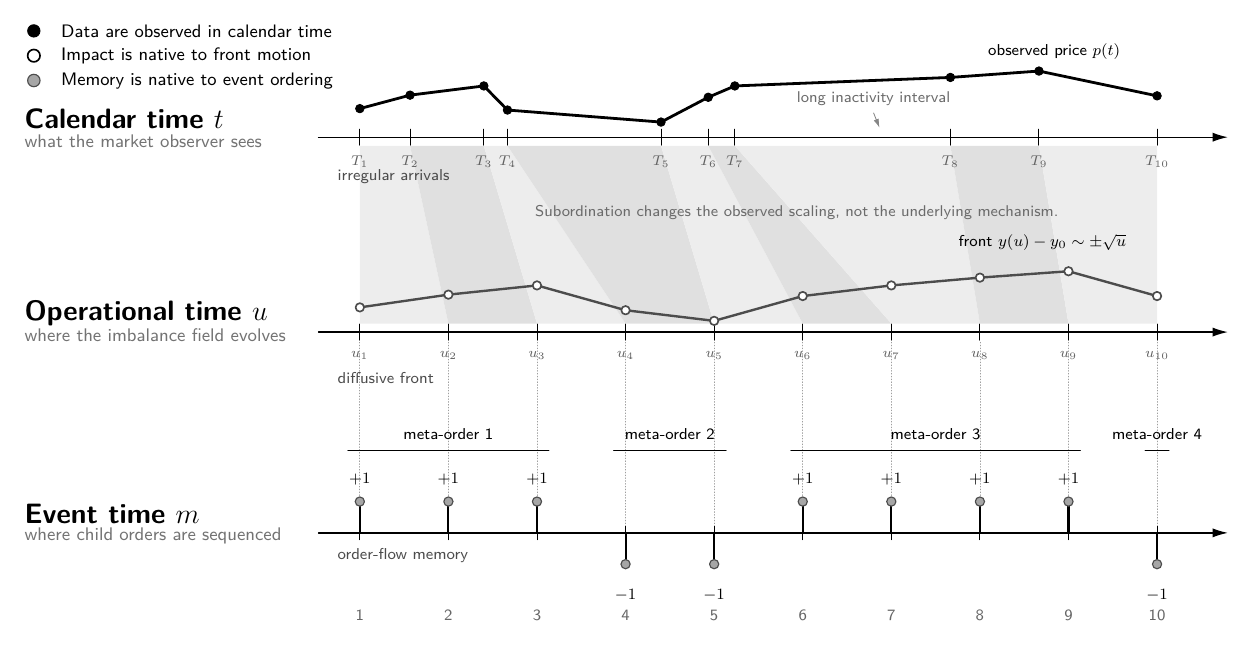}
\caption{\textbf{Schematic separation of event, operational, and calendar time.}
The lower layer shows an illustrative sequence of signed child orders indexed by event number $m$, grouped into same-sign meta-orders, corresponding to the event-time renewal mechanism described in Section~\ref{sec:event-renewal}. These events define the regular operational mesh $u_m=m\Delta u$ of Section~\ref{sec:operational-time}, on which the imbalance field and its reaction front evolve according to Eq.~\eqref{eq:rapid_pde}. The drawn front segments schematically illustrate the signed square-root response obtained from the Volterra front equation~\eqref{eq:rapid_volterra} in the constant-rate, locally linear, frozen-background, weak-cancellation/small-displacement regime, giving Eq.~\eqref{eq:rapid_sqrt}; they are not intended as an exact solution for the concatenated sequence. The upper layer maps the corresponding operational observations to irregular calendar arrival times, as described by the subordination construction in Section~\ref{sec:calendar-time}, with the observed front represented by the clock-time impact process~\eqref{eq:rapid_clock_impact}. Black points mark calendar-time observations; the connecting line is graphical interpolation only and does not imply price evolution during intervals with no events.}
\label{fig:clock-separation}
\end{figure*}

This gives the minimal bookkeeping needed to keep these regularities distinct while deriving them from a common bid/ask order-book representation. The steps are: (i) an exact event-time imbalance reduction; (ii) an operational-time locally linear latent-order-book approximation; and (iii) a separate stochastic event clock that maps event or operational time into calendar time. This leaves three times in play: the event index $m$ native to child orders, the operational continuum time $u$ associated with the reaction--diffusion field, and the calendar time $t$ generated by event waiting times. Figure~\ref{fig:clock-separation} provides a schematic summary of this clock hierarchy.

\section{Discrete Event-Time Order Book and Exact Imbalance Reduction}

Let $\rho_{B,j}^m$ and $\rho_{A,j}^m$ denote the bid- and ask-side densities at log-price lattice site $x_j=j\Delta x$ immediately after event $m$. Between events, liquidity diffuses, cancels, is replenished by lit and latent sources, and is perturbed by signed child-order forcing. 

Given a survival function $\theta^{n,m-1}_j$ from event $n$ to $m-1$ and a memory kernel $K_{m-n}$, we can define a discrete time random walk model on a random lattice with diffusion intensities given in terms of left, right and self jump probabilities \cite{angstmann2015dtrw,diana2024anomalous}
\begin{equation}
    \lambda_{j|j+i}^{m} = \tfrac{r}{2} \delta_{j-1,j+i}+ (1-r) \delta_{j,j+i}+\tfrac{r}{2} \delta_{j+1,j+i}. \label{eq:TransitionProb1}
\end{equation}

The generic symmetric bid/ask update then follows from combining the anomalous diffusions with orders that have survived, symmetric reactions across the spread, sources and meta-order forcing at the front:
\begin{align}
\rho_{B,j}^{m}
&=\sum_{i=-1}^{1}\lambda_{j|j+i}^{m}
  \sum_{n=0}^{m-1}K_{m-n}\theta_{j+i}^{n,m-1}
  \rho_{B,j+i}^{n}
\nonumber\\
&\quad+\theta_{j}^{m,m-1}\rho_{B,j}^{m-1}
 -\sum_{n=0}^{m-1}K_{m-n}\theta_{j}^{n,m-1}
  \rho_{B,j}^{n}
\nonumber\\
&\quad-\kappa\tau_m\rho_{A,j}^{m-1}\rho_{B,j}^{m-1}
 +\tau_mS_{j,m}^{B}+M_{j,m}^{B},
\label{eq:rapid_bid_full}
\\[2pt]
\rho_{A,j}^{m}
&=\sum_{i=-1}^{1}\lambda_{j|j+i}^{m}
  \sum_{n=0}^{m-1}K_{m-n}\theta_{j+i}^{n,m-1}
  \rho_{A,j+i}^{n}
\nonumber\\
&\quad+\theta_{j}^{m,m-1}\rho_{A,j}^{m-1}
 -\sum_{n=0}^{m-1}K_{m-n}\theta_{j}^{n,m-1}
  \rho_{A,j}^{n}
\nonumber\\
&\quad-\kappa\tau_m\rho_{B,j}^{m-1}\rho_{A,j}^{m-1}
 +\tau_mS_{j,m}^{A}+M_{j,m}^{A}.
\label{eq:rapid_ask_full}
\end{align}
Here $\tau_m$ is the waiting time between event $m-1$ and event $m$, the sources are $S$, the meta-orders forcing the front are $M$, $\nu$ is a cancellation rate, and $\kappa$ the reaction rate. The full form is included to make the clock dependence explicit; the remainder of the paper uses the Markovian nearest-neighbour reduction.

Assume exponential arrivals $\theta_j^{m,m-1}=e^{-\nu\tau_m}$ and no memory, $K_{m-n}=\delta_{n,m-1}$. Taking the diffusive limit $D=\lim_{\Delta x,\tau_m\to0}(r/2)\,\Delta x^2/\tau_m$, the bid/ask updates become
\begin{align}
\rho_{B,j}^{m}
&= \rho_{B,j}^{m-1}
 + \frac{D\tau_m}{(\Delta x)^2}
   \bigl(\rho_{B,j+1}^{m-1}-2\rho_{B,j}^{m-1}
   +\rho_{B,j-1}^{m-1}\bigr)
\nonumber\\
&\quad -\nu\tau_m\rho_{B,j}^{m-1}
 -\kappa\tau_m\rho_{A,j}^{m-1}\rho_{B,j}^{m-1}
\nonumber\\
&\quad +\tau_m S_{j,m}^{B}+M_{j,m}^{B},
\label{eq:rapid_bid}
\\[2pt]
\rho_{A,j}^{m}
&= \rho_{A,j}^{m-1}
 + \frac{D\tau_m}{(\Delta x)^2}
   \bigl(\rho_{A,j+1}^{m-1}-2\rho_{A,j}^{m-1}
   +\rho_{A,j-1}^{m-1}\bigr)
\nonumber\\
&\quad -\nu\tau_m\rho_{A,j}^{m-1}
 -\kappa\tau_m\rho_{A,j}^{m-1}\rho_{B,j}^{m-1}
\nonumber\\
&\quad +\tau_m S_{j,m}^{A}+M_{j,m}^{A}.
\label{eq:rapid_ask}
\end{align}
Define the event-time imbalance field
\begin{equation}
\phi_j^m:=\rho_{B,j}^m-\rho_{A,j}^m.
\end{equation}
Subtracting Eq.~\eqref{eq:rapid_ask} from Eq.~\eqref{eq:rapid_bid} eliminates the symmetric matching term exactly and yields the closed recursion
\begin{align}
\phi_j^m
&=\phi_j^{m-1}
 +\frac{D\tau_m}{(\Delta x)^2}
  \bigl(\phi_{j+1}^{m-1}-2\phi_j^{m-1}
  +\phi_{j-1}^{m-1}\bigr)
\nonumber\\
&\quad -\nu\tau_m\phi_j^{m-1}
 +\tau_m s_{j,m}+\mathcal M_{j,m},
\label{eq:rapid_imbalance}
\end{align}
with $s_{j,m}=S_{j,m}^{B}-S_{j,m}^{A}$ and $\mathcal M_{j,m}=M_{j,m}^{B}-M_{j,m}^{A}$. The cancellation in the imbalance equation is exact conditional on the realised waiting-time sequence and it does not depend on the law of the waiting times. The subsequent continuum operational-time limit is a separate scaling step. This links the bid/ask system to the mesoscopic front dynamics.

The discrete mid-price proxy is the zero of the imbalance field and occurs at grid index $j^*_m$. If $j_m^*$ satisfies $\phi_{j_m^*}^m\phi_{j_m^*+1}^m\le 0$, then a front estimate is obtained by linear interpolation,
\begin{equation}
p_m=x_{j_m^*}+\Delta x\left(
\frac{-\phi_{j_m^*}^m}{\phi_{j_m^*+1}^m-\phi_{j_m^*}^m}
\right).
\end{equation}

\section{Operational-Time Continuum and Square-Root Impact}
\label{sec:operational-time}

To parametrise the continuum event-time limit, introduce the interpolant $\Phi^{\Delta x,\Delta u}$ of the discrete imbalance field, defined by $\Phi^{\Delta x,\Delta u}(x_j,u_m)=\phi_j^m$ \cite{angstmann2015dtrw}. Here $u_m=m\Delta u$ is the operational-time mesh. In the passage to \(u\), the random calendar waiting times are not being sent to zero; they are held for the later subordination step. The limit is an operational/event-count scaling of the imbalance field. 

Then passing formally to a continuum operational clock $u$, the imbalance field $\Phi(x,u)$ satisfies
\begin{align}
\partial_u\Phi(x,u)
&=D_u\partial_{xx}\Phi(x,u)-\nu_u\Phi(x,u)
\nonumber\\
&\quad+s_u(x,y(u))+m_u(x,u),
\label{eq:rapid_pde}
\end{align}
where $y(u)$ is the operational-time reaction front defined by $\Phi(y(u),u)=0$. The stationary background $\Phi^*(x)$ solves the corresponding source-balance equation with $m_u\equiv 0$. Near its simple zero $y_0$, the stationary profile is locally linear,
$\Phi^*(x)\approx-\mathcal L_u(x-y_0)$, 
with local liquidity slope $\mathcal L_u=-\partial_x\Phi^*(y_0)>0$. This is the locally linear latent-order-book approximation \cite{donier2015fully,mastromatteo2014latent,benzaquen2018multi}.

Writing $\Phi(x,u)=\Phi^*(x)+\Psi(x,u)$ and freezing source adaptation over the execution horizon leads to the perturbation equation
\begin{align}
\partial_u\Psi(x,u)
&=D_u\partial_{xx}\Psi(x,u)-\nu_u\Psi(x,u)
\nonumber\\
&\quad+m_e(u)\delta(x-y(u)).
\end{align}
with $\Psi(x,0)=0$. Using the full-line damped diffusion kernel and evaluating the perturbation at the moving front gives the operational-time Volterra equation, in the locally linear latent-order-book form of \cite{donier2015fully} (see also \cite{toth2011anomalous,mastromatteo2014latent,benzaquen2018multi}):
\begin{align}
\mathcal L_u\bigl[y(u)-y_0\bigr]
&=\int_0^u m_e(s)
  \frac{e^{-\nu_u(u-s)}}{\sqrt{4\pi D_u(u-s)}}
\nonumber\\
&\qquad{}\times
  \exp\!\left[-\frac{(y(u)-y(s))^2}{4D_u(u-s)}\right]ds.
\label{eq:rapid_volterra}
\end{align}

For constant-rate execution $m_e(s)=m_0\mathbf{1}_{[0,U]}(s)$ and in the weak-cancellation, small-displacement regime, Eq.~\eqref{eq:rapid_volterra} reduces to the Abel-kernel form of the locally linear latent-order-book impact equation \cite{donier2015fully,mastromatteo2014latent,toth2011anomalous}, and yields the operational-time square-root law
\begin{equation}
y(u)-y_0\approx \frac{m_0}{\mathcal L_u}\sqrt{\frac{u}{\pi D_u}},
\qquad 0\le u\le U.
\label{eq:rapid_sqrt}
\end{equation}
At completion $Q_U=m_0U$, the impact is therefore proportional to $\sqrt{Q_U}$ at fixed operational-time participation rate. The square-root law is generated by diffusive memory on the operational clock, not by the event-ordering mechanism that produces sign persistence.

\section{Event-Time Renewal Law for Trade Signs}
\label{sec:event-renewal}
The child-order sign process $\epsilon_m\in\{\pm1\}$ is defined on the event clock. In the sequential hidden-order order-splitting picture \cite{lillo2005theory}, meta-orders have i.i.d. signs and i.i.d. lengths $L$ with law $p_L(\ell)=\Pp[L=\ell]$. A uniformly sampled child-order does not see $p_L$, but the length-biased law
\begin{equation}
\widetilde p_L(\ell)=\frac{\ell p_L(\ell)}{\E[L]}.
\end{equation}
The event-time sign autocorrelation is then exactly
\begin{equation}
C_{\tau}(\epsilon)
=\frac{1}{\E[L]}\sum_{\ell=\tau+1}^{\infty}(\ell-\tau)p_L(\ell).
\label{eq:rapid_lmf_exact}
\end{equation}
If $p_L(\ell)\sim c_L\ell^{-(\alpha+1)}$, then
\begin{equation}
C_{\tau}(\epsilon)\sim K\tau^{-(\alpha-1)},
\qquad \gamma=\alpha-1.
\label{eq:rapid_lmf_asymptotic}
\end{equation}
This is the Lillo--Mike--Farmer exponent relation \cite{lillo2005theory,sato2023prl,sato2024jsp} and the mechanism is purely renewal-theoretic and depends only on event ordering and hidden-order lengths.

The event-time sign process also admits a natural source-based interpretation in the reaction--diffusion model. If the net forcing entering the imbalance equation factorizes locally as $\mathcal M_{j,m}=\sigma_m q_{j,m}$ with $q_{j,m}\ge 0$ supported near the front, then the aggregate local forcing has the same sign as the primary trade sign. The hidden-order sign law can therefore be read as a law for persistent interface forcing. This identification is exact once the local forcing factorisation is imposed, and does not require a statement about prices or market efficiency. Other conventions for assigning trade signs, such as flux-based or interface-crossing definitions, change the representation of the sign observable but not the clock distinction: the persistence law remains indexed by event ordering before calendar-time sampling is applied \cite{lillo2005theory,sato2023prl,sato2024jsp}.

\section{Calendar-Time Sampling and Subordination}
\label{sec:calendar-time}
Let the cumulative waiting-time process be $T_m=\sum_{k=1}^{m}\tau_k$ and the inverse event counter be
\begin{equation}
N_t=\max\{m\ge0:T_m\le t\}.
\end{equation}
The sampled sign process in calendar time is then $\epsilon(t)=\epsilon_{N_t}$. For a renewal clock started at an event epoch, the clock-time correlation can be written as
\begin{equation}
C_t(\epsilon)=\sum_{\tau=0}^{\infty}\Pp(N_t-N_0=\tau)C_{\tau}(\epsilon).
\label{eq:rapid_clock_sign}
\end{equation}
The point being made here is that the event-time sign law is observed only after composition with the event clock. Similarly, the observed front is the subordinated process
\begin{equation}
p(t)=y(E_t),
\end{equation}
Here $E_t$ is the inverse subordinator in the continuum scaling limit \cite{angstmann2015dtrw,meerschaert2012fractional,diana2024anomalous}. The execution schedule is specified on the operational clock. A calendar-time execution schedule would require subordinating the forcing process as well as the front. Substituting Eq.~\eqref{eq:rapid_sqrt} gives the clock-time impact process conditional on the realised inverse clock,
\begin{equation}
I(t):=p(t)-y_0\approx \frac{m_0}{\mathcal L_u}\sqrt{\frac{E_t}{\pi D_u}}.
\label{eq:rapid_clock_impact}
\end{equation}
If the waiting-time law has finite mean, then the inverse clock grows linearly on average, $E_t$ is of order $t/\bar\tau$, and the standard clock-time square-root law is recovered up to a deterministic activity rescaling. If the waiting-time law is heavy-tailed with stable index $\beta\in(0,1)$, then the inverse clock has typical and moment scaling of order $t^\beta$, and the corresponding impact scale behaves as $t^{\beta/2}$. The operational-time square-root mechanism survives intact; only the projection from operational time to calendar time changes.

Fractional time enters the square-root law only through the event-to-calendar projection. The Abel-kernel/front-motion mechanism remains operational-time. Fractional or tempered behaviour enters through the event-to-calendar projection: stable clocks give an anomalous \(t^{\beta/2}\)-type impact scale, while tempered clocks cross over to ordinary finite-mean scaling at long horizons.

\section{Discussion}
The construction here is a clock-separation result. The separate ingredients are familiar: order splitting gives persistent order-flow signs, reaction--diffusion latent-liquidity models give concave impact, and random waiting times give calendar-time sampling effects. What is less often kept explicit is that these are not native to the same clock. Placing them on one time line implicitly chooses a subordination, and this changes the interpretation of both measured memory and measured impact.

It is useful to distinguish this construction from renewal-theoretic perspectives \citea{sato2023prl,sato2024jsp}. That work gives a quantitative and exact treatment of the Lillo--Mike--Farmer mechanism, showing how persistent market-order signs can be inferred from order-splitting structure and heterogeneous meta-order lengths \cite{sato2023prl,sato2024jsp,lillo2005theory}. The present argument does not modify that event-time mechanism. Rather, it fixes its clock status inside a bid/ask reaction--diffusion representation: the LMF law remains a statement about the ordering of child orders, while price impact is produced by the evolution of an imbalance field and its reaction front. In this sense the renewal law is used here as an event-time source law for persistent interface forcing, not as a direct calendar-time price-impact law. The relation to the latent-liquidity work is similarly complementary \citea{toth2011anomalous,mastromatteo2014latent,donier2015fully,benzaquen2018multi,bouchaud2018tqp}. The locally linear order-book approximation, the reaction--diffusion front equation, and the resulting square-root impact law follow the same structural route developed in this literature \cite{toth2011anomalous,mastromatteo2014latent,donier2015fully,benzaquen2018multi}. 

Related information-processing accounts also recover square-root impact and reversion effects \citea{saddier2024bayesian}. The distinction here is that impact is first formulated on an operational clock and only then sampled in calendar time. Thus clock-time impact can change with the sampling and waiting-time process even when the underlying latent-liquidity square-root mechanism is unchanged.

This clock separation also changes how universality claims should be read. The empirical or model-level observation of a square-root law is not, by itself, a statement about the uniqueness of the time variable on which the law is formulated. In the Sato--Kanazawa approach, the event sequence is the natural object: account-level order splitting supports the microscopic source of persistent order-flow signs, and recent empirical and solvable constructions relate long-range order-flow correlation to square-root impact \cite{sato2023prr,sato2025prl,sato2025solvable}. In the latent-liquidity approach, by contrast, the square-root law is produced by reaction--diffusion front motion in a locally linear hidden order book \cite{donier2015fully,mastromatteo2014latent,benzaquen2018multi}. The present construction is distinct to both: it asks whether the renewal law, the front law, and the measured calendar-time observable are being compared on the same clock or only after an implicit subordination. Thus an apparently universal law may be stable as an event-time or operational-time statement while changing its empirical meaning under a different event-to-calendar mapping.

The role of tail truncation, finite-size effects, heterogeneity, or other regularisations inside event-time order-splitting models is different from the role played by subordination here. In the Sato--Kanazawa line of attack, these ingredients concern the event-indexed source mechanism and, in recent solvable work, its connection to nonlinear impact through a L\'evy-walk representation \cite{sato2025solvable}. In contrast, following the present DTRW construction, fractional or tempered behaviour enters through the clock projection from event or operational time to calendar time. It is therefore not only a modification of the source-process tail, but part of the definition of the observed market process.

This clarifies why the shape of price impact can depend on the discretisation procedure used in simulation work \cite{diana2024anomalous}: the discretisation carries an implicit subordination. It also clarifies part of the reason why persistent order flow and weak return autocorrelation can coexist. Long-memory signs do not need to imply long-memory returns because impact is filtered by transient response, replenishment, cancellation, and local liquidity adaptation before it is sampled on the physical clock \cite{bouchaud2004fluctuations,bouchaud2006liquidity,gatheral2010nda,farmer2013impact}. This is consistent with the adaptive-liquidity view in which predictable order-flow components are partly absorbed by state-dependent liquidity taking and order-book response \cite{taranto2014adaptive}. 

A persistent source therefore need not generate persistent returns; the return is not the source itself, but the filtered and subordinated displacement of the reaction front \cite{bouchaud2004fluctuations,taranto2014adaptive}. There are therefore at least two distinct filtering layers: first, from event-time forcing to operational-time front motion; second, from operational-time observables to calendar-time measurements.

\section{Conclusion}
In summary, the impact calculation assumes a locally linear latent order book and a frozen-background execution regime. The event clock is exogenous rather than state dependent. These restrictions are useful because they make clear which statements are exact in the event-indexed bid/ask reduction, which belong to operational-time transport, and which arise only after subordination.

The practical consequence is that the fitted object depends on the clock. A calendar-time impact curve is not a direct estimate of the operational front law; it is the front law after the execution schedule, event intensity, waiting-time process, and interpolation rule have been composed into the observable. This matters for trading costs, because square-root coefficients estimated across assets or venues mix liquidity response with clock projection unless the sampling clock is fixed. The same warning applies elsewhere: once the clock is changed, the fitted statistic is not the primitive mechanism \cite{angstmann2026reactionboundary}.

\section*{Acknowledgments}

We are thankful for the input of Derick Diana, Byron Jacobs and Dominic Bauer. We thank the reviewers for thoughtful and helpful feedback.

\section*{Conflict of interest}
The authors declare no competing interests.

\section*{AI disclosure}
The authors used generative AI tools for language and formatting checks. The scientific content and final text were controlled by the authors.

\bibliographystyle{elsarticle-num}
\bibliography{CATG-ClockSeparation-v1.0.2}
\end{document}